\documentclass[conference]{IEEEtran}
\IEEEoverridecommandlockouts
\usepackage{cite}
\usepackage{amsmath,amssymb,amsfonts}
\usepackage{algorithmic}
\usepackage{graphicx}
\usepackage{subcaption}
\usepackage{float}
\usepackage{textcomp}
\usepackage{xcolor}
\usepackage{url}
\usepackage{multirow}
\usepackage{booktabs} 
\usepackage[acronym]{glossaries}

\newacronym{cad}{CAD}{computer-aided diagnosis}
\newacronym{gan}{GAN}{generative adversarial network}
\newacronym{ml}{ML}{machine learning}
\newacronym{ssim}{SSIM}{structural similarity index}
\newacronym{psnr}{PSNR}{peak signal-to-noise ratio}
\newacronym{fid}{FID}{Fréchet inception distance}
\newacronym{gi}{GI}{gastrointestinal}

\def\BibTeX{{\rm B\kern-.05em{\sc i\kern-.025em b}\kern-.08em
    T\kern-.1667em\lower.7ex\hbox{E}\kern-.125emX}}
\begin{document}

\title{PolypConnect: Image inpainting for generating realistic gastrointestinal tract images with polyps
}

% Authors

% Jan Andre Fagereng
% Vajira Thambawita
% Andrea
% Thomas
% Sravanthi
% Pål
% Michael

\author{\IEEEauthorblockN{Jan Andre Fagereng}
\IEEEauthorblockA{\textit{Oslo Metropolitan University} \\
Oslo, Norway \\
andre\_fagereng@hotmail.com}
\and
\IEEEauthorblockN{Vajira Thambawita}
\IEEEauthorblockA{\textit{SimulaMet} \\
Oslo, Norway \\
vajira@simula.no}
\and
\IEEEauthorblockN{Andrea M. Stor{\aa}s}
\IEEEauthorblockA{\textit{SimulaMet} \\
Oslo, Norway \\
andrea@simula.no}
\and
\IEEEauthorblockN{Sravanthi Parasa}
%\IEEEauthorblockA{\textit{Department of Gastroenterology} \\
\IEEEauthorblockA{\textit{Swedish Medical Group} \\
%\textit{Swedish Medical Group}\\
Seattle, WA, USA \\
vaidhya209@gmail.com}
\and 
\IEEEauthorblockN{Thomas de Lange}
%\IEEEauthorblockA{\textit{Department of Medicine and Emergencies,}\\
\IEEEauthorblockA{\textit{Sahlgrenska University Hospital, Gothenburg}\\
V\"{a}stra G\"{o}taland Region, Sweden \\
thomas.de.lange@gu.se}
\and
\IEEEauthorblockN{P{\aa}l Halvorsen}
\IEEEauthorblockA{\textit{SimulaMet} \\
Oslo, Norway \\
paalh@simula.no}
\and
\IEEEauthorblockN{Michael A. Riegler}
\IEEEauthorblockA{\textit{SimulaMet} \\
Oslo, Norway \\
michael@simula.no}

}

\maketitle

%\vspace{-10pt}
\begin{abstract}

Early identification of a polyp in the lower gastrointestinal (GI) tract can lead to prevention of life-threatening colorectal cancer. Developing computer-aided diagnosis (CAD) systems to detect polyps can improve detection accuracy and efficiency and save the time of the domain experts called endoscopists. Lack of annotated data is a common challenge when building CAD systems. Generating synthetic medical data is an active research area to overcome the problem of having relatively few true positive cases in the medical domain. To be able to efficiently train machine learning (ML) models, which are the core of CAD systems, a considerable amount of data should be used. In this respect, we propose the PolypConnect pipeline, which can convert non-polyp images into polyp images to increase the size of training datasets for training. We present the whole pipeline with quantitative and qualitative evaluations involving endoscopists. The polyp segmentation model trained using synthetic data, and real data shows a $5.1\%$ improvement of mean intersection over union (mIOU), compared to the model trained only using real data. The codes of all the experiments are available on GitHub to reproduce the results. 
\end{abstract}


\section{Introduction}
Utilizing the potential of data and deep learning in the medical sphere is a highly regarded and valuable task. Intelligent tools and \gls{cad} systems \cite{cad_1, cad_2, cad_3} can be developed in order to assist medical staff, in an effort to increase precision in diagnosis, support or guide in decision-making, or increase the general efficiency of medical processes. Even though there are clear potentials in utilizing artificial intelligence for such tasks, several challenges still exist to be researched.

One of the major issues in developing robust tools utilizing \gls{ml} algorithms within the medical sphere is the lack of annotated data. Manual annotation of data by domain experts is a costly and time-consuming process, which is impractical in order to generate a substantially sized dataset for model consumption. Moreover, the manual data annotation process is subjective. As \gls{cad} systems potentially have an impact on the actions or decisions of doctors and medical employees, it is crucial to obtain robust and reliable models. Models trained on small datasets might yield predictions with over-fit assumptions, not suitable for out-of-sample data and unfit for a production setting. The use of sensitive patient data can also give rise to privacy-related issues, which complicates the open sharing of data and code.

In this research, we aim to reduce or circumvent the issues above by producing machine-generated synthetic images with respective annotations in a selected medical case study, namely polyp segmentation~\cite{polyp_seg_1, polyp_seg_2, polyp_seg_3}. More specifically, we apply image inpainting to generate \gls{gi} images containing polyps. Image inpainting can be described as a method that estimates pixel values to fill holes or missing areas in an image. By utilizing both unlabeled and labeled data, we train three image inpainting models and analyze the performance of generating polyps on clean-colon images. This is a suitable method since the mucosa surrounding the polyp is mostly completely normal. Finally, we introduce an effective polyp inpainting pipeline, called \textbf{PolypConnect}, to generate synthetic polyps in clean colon data based on the best findings of our experiments. This generation process is an effort to enlarge the size of the dataset and subsequently compare segmentation models trained on a mix of real and synthetic images to evaluate performance. The goal is to research if a generation of realistic images is viable for this kind of data and to what extent it has an impact on improving polyp segmentation models. Moreover, since the generated polyp images are not from real patients, this could be a way of circumventing regulations related to the privacy protection of sensitive medical data and sharing data more easily.

In this regard, we can list our main contributions as follows:
% Write here
\begin{itemize}
    \item We evaluate three different state-of-the-art image inpainting models for the \gls{gi} tract and benchmark the best model in our polyp inpainting pipeline. 
    \item We introduce PolypConnect, an efficient (in terms of usability)  polyp inpainting pipeline to convert non-polyp images (true-negative samples) to realistic polyp images (true-positive samples).
    \item We evaluate the quality of the pipeline quantitatively and qualitatively with the aid of medical experts. 
    \item We evaluate the effectiveness of using synthetic polyp data for polyp segmentation models using the UNet architecture.  
\end{itemize}

The code is available in \url{https://github.com/AndreFagereng/polyp-GAN} to reproduce the results and future studies, and this work is building upon the work by Thambawita et al.~\cite{thambawita2021id}.

\section{Related work}

Generating synthetic polyps is not a new research direction. However, producing realistic synthetic polyps with the corresponding ground truth, which can be used to train other machine learning models, is challenging. Random synthetic \gls{gi}-tract images can be generated from the pre-trained \gls{gan} models studied in ~\cite{deepsynthbody, yoon2022colonoscopic}. However, generating synthetic polyps and corresponding ground truth is not possible with these model. 

One study developed an inpainting method for endoscopy medical images which was also able to remove specular highlights of polyps~\cite{arnold2010polypinpainting}. Akbari et al.~\cite{akbari2018inpainting} removed reflections in colonoscopy video frames using a proposed inpainting method. A \gls{gan} has also been developed to do inpainting reflections in endoscopic images~\cite{funke2018GANinpainting}. Recently, Daher et al. developed a temporal \gls{gan} for the same purpose~\cite{daher2022GANinpainting}. However, none of these methods are designed to inpaint synthetic polyps in clean \gls{gi} images.

The SinGAN-Seg pipeline was introduced by Thambawita et al.~\cite{singan-seg} to generate synthetic polyps with the corresponding segmentation masks. Because this model uses only a single real polyp as an input, the generated samples show very close distributions of pixels to the input image. Furthermore, this model was developed to generate completely new synthetic polyp images from scratch and was not tested for converting non-polyp images into polyp images. Also, distributions of synthetic polyp images generated from this kind of pure polyp generators are showing quite similar distribution to the training data used to train~\glspl{gan} of the pipelines.   

A simple UNET-based synthetic polyp generator was introduced by Qadir et al.\cite{qadir2022simple}. In this study, they have experimented with converting polyp images into non-polyp and non-polyp into polyp images. However, the generated polyp can be discriminated easily based on the presented results. Furthermore, using only the mask to generate polyps makes generated polyps unrealistic, and the structure of the polyp can not be adjusted. In this regard, we present the PolypConnect pipeline to generate realistic synthetic polyp on clean colon images. 

\section{PolypConnect pipeline}

\begin{figure*}[htbp]
\centerline{\includegraphics[width=\textwidth]{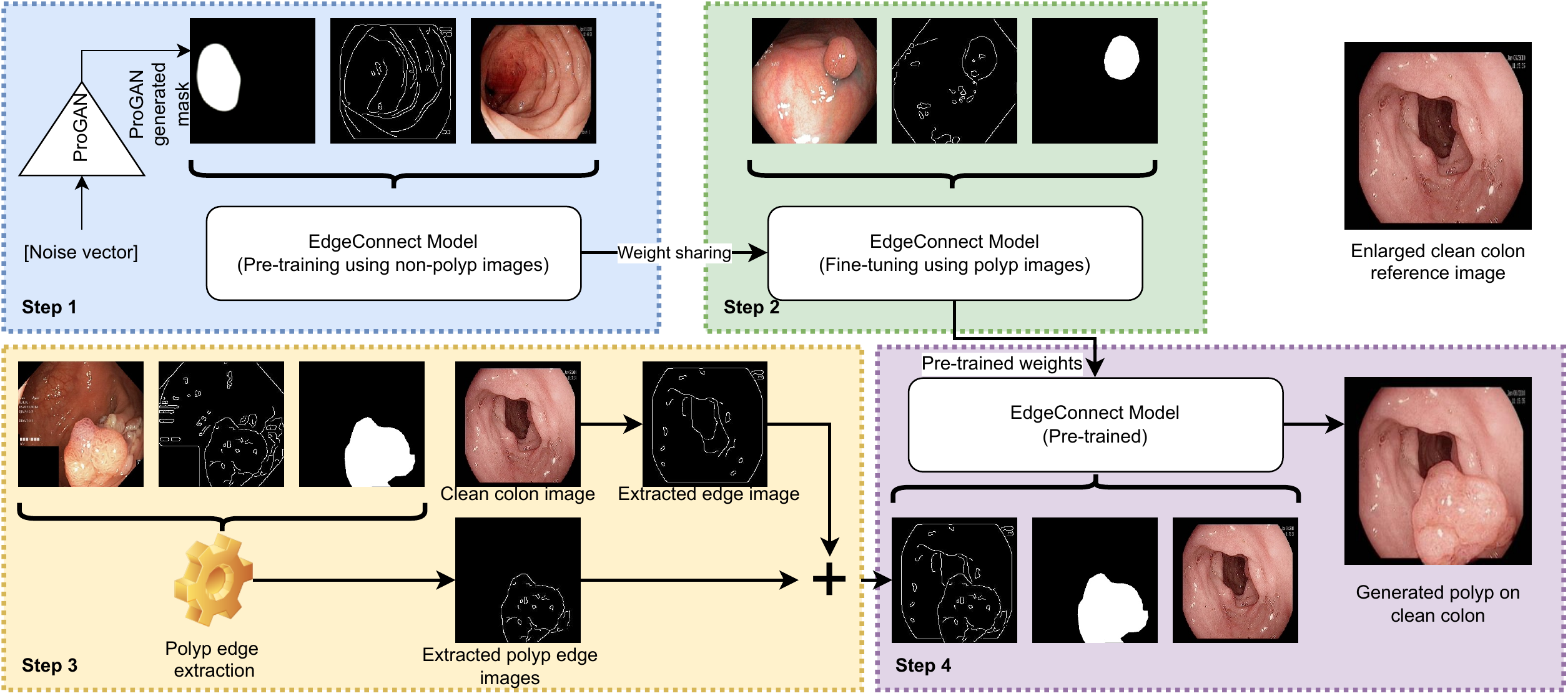}}
\caption{Pipeline of PolypConnect. \textbf{Step 1} - Pre-training of EdgeConnect , \textbf{Step 2} - Fine-tuning of the EdgeConnect model, \textbf{Step 3} - Edge extractions for polyp masks (an alternative method is discussed in Section~\ref{sec:discussion}), \textbf{Step 4} - Generating polyps on clean colon images.}
\label{fig:pipe_line}
\vspace{-10pt}
\end{figure*}

The complete pipeline of PolypConnect is depicted in Figure \ref{fig:pipe_line}. The pipeline consists of four steps. In \textbf{Step 1}, we use a \gls{gan} architecture to generate synthetic realistic polyp masks. In this study, we have used ProGAN~\cite{karras2017progressive}. However, other \gls{gan} architectures such as StyleGAN~\cite{karras2019style} and FastGAN~\cite{liu2020towards} can be used for the synthetic mask generation. The generated synthetic masks are then randomly paired with \gls{gi}-tract images to produce images with missing regions. The EdgeConnect model is then pre-trained, conditioned on the missing region images and extracted edge maps. 

In \textbf{Step 2}, we fine-tune the pre-trained EdgeConnect model using polyp datasets with corresponding extracted edge images. The pre-trained weights of the EdgeConnect model were loaded from Step 1. In this process, the model is trained to inpainting the exact polyp regions using the manually annotated ground truth provided in the datasets.  

\textbf{Step 3} prepares the input data in order to produce realistic polyp output from non-polyp colon images. As a simple method, we extract polyp masks and the corresponding edge from the polyp datasets used in Step 2. Alternatively, polyp edge and corresponding masks can be generated using another \gls{gan} model. Then, extracted polyp edge is merged into the edge image of a clean colon image. 

In \textbf{Step 4}, the edge polyp image returned from Step 3, the corresponding mask, and the clean colon image are provided as input to the pre-trained EdgeConnect model of Step 2. The generated polyp image is the final output of this PolypConnect pipeline. A sample clean-colon image and the generated polyp image generated using it are depicted in Figure~\ref{fig:pipe_line}.

\section{Experimental results}

\subsection{Data}
For the purpose of training the models for the generation of synthetic polyps, we used the HyperKvasir dataset \cite{borgli2020hyperkvasir}. It consists of $1,000$ images with segmented polyp annotations and around $100,000$ unlabeled \gls{gi}-tract landmarks. In Step 1, we have used the unlabeled data to pre-train the models. The segmentation dataset was used from Step 2 to the final polyp segmentation experiments. For the validation procedure, $200$ images were randomly sampled from the training data, and therefore the models were trained with the remaining 800 images. This initial split $(80/20)$ of the dataset was kept throughout all of the experiments, including the segmentation performed at the end. All the data used in this study are anonymous and publicly available for research purposes.

\begin{figure}
    \centering
    \setlength{\tabcolsep}{1pt}
    \newcommand{\imgwidth}{0.22\linewidth}
    \begin{tabular}{cccc}
    
    \includegraphics[width=\imgwidth]{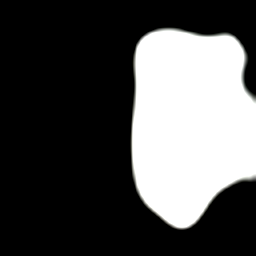} &
    \includegraphics[width=\imgwidth]{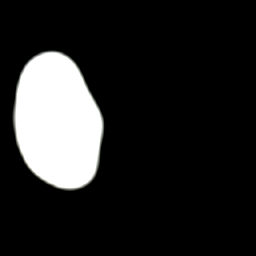} &
    \includegraphics[width=\imgwidth]{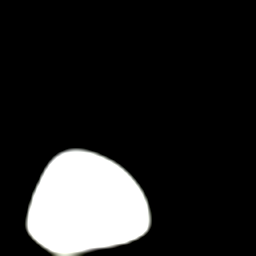} &
    \includegraphics[width=\imgwidth]{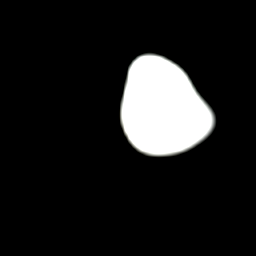}
    \\
  
    \includegraphics[width=\imgwidth]{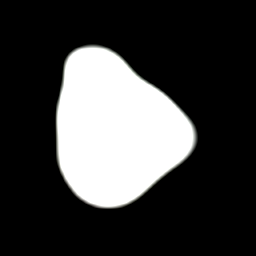} &
    \includegraphics[width=\imgwidth]{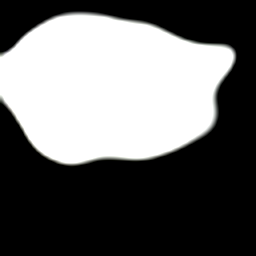} &
    \includegraphics[width=\imgwidth]{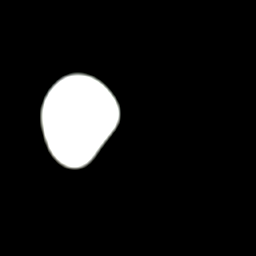} &
    \includegraphics[width=\imgwidth]{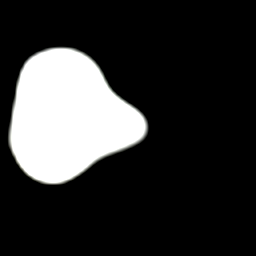}

    \\

     \end{tabular}
    
    \caption{Sample synthetic masks generated from ProGAN~\cite{karras2017progressive}}
    \label{fig:synthetic_masks}
    \vspace{-12pt}
\end{figure}

\subsection{Experimental setup}
We used two GPUs of Geforce GTX TITAN X of $8$ GB for the preliminary experiments. When the experiments were ready to run for a total number of epochs, an NVIDIA DGX server having $16$ v100 GPUs of $32$ GB was used, and the Pytorch deep learning framework version 1.10.1 was used for all experiments. 

The \gls{ssim}~\cite{ssim}, \gls{psnr}~\cite{korhonen2012peak} and \gls{fid}~\cite{heusel2017fid} were used for quantitative evaluation of the models on the validation data. These metrics are commonly used to evaluate inpainting methods. Ideally, the \gls{ssim} and \gls{psnr} should be as high as possible, while the \gls{fid} should be close to $0$. In addition, a user survey including medical doctors was performed in order to evaluate the models qualitatively. Details about the survey are provided in subsection~\ref{sec:expert_survey}.

\subsection{Synthetic polyp masks}
The total number of unique segmented polyp masks from KvasirSEG\cite{jha2020kvasir} is $1,000$, which is insufficient for our preliminary experiments of general image inpainting. Thus, we used ProGAN~\cite{karras2017progressive} conditioned on random noise to produce new realistic polyp masks. Figure~\ref{fig:synthetic_masks} visualizes some samples of the generated masks. Examining the generated masks, we observed a couple of undesirable shapes and sizes. The mask regions in these images were either taking up the entire region or were non-existent. In other words, the binary distribution of the masks was at each end of the extremes in these cases. Therefore, we decided to discard such images by only keeping generated images where the masked regions filled between $5\%-70\%$ of the entire image.

\subsection{Polyp inpainting}

We have selected three image inpainting models, namely GMCNN~\cite{gmcnn}, AOTGAN~\cite{aotgan} and EdgeConnect~\cite{edgeconnect}, to explore the capability of inpainting polyps on a given clean colon image. These three models were chosen due to their popularity and novelty. We have performed a set of preliminary experiments, which are presented in Table~\ref{tab:compare_models}. Based on the preliminary results, we chose to proceed with EdgeConnect and AOTGAN.
\begin{table}[]
    \centering
    \caption{Comparison of different image inpainting models. Selected best values from different checkpoints are presented here. The best two models' values are presented using \textbf{bold} text.}
    \begin{tabular}{|c|c|c|c|}
    \toprule
        Model & SSIM & PSNR & FID  \\
    \midrule
         GMCNN~\cite{gmcnn} & 0.4911 & 12.641 & 181.720 \\
         EdgeConnect~\cite{edgeconnect} & \textbf{0.6100} & \textbf{17.980} & \textbf{74.070} \\
         AOTGAN~\cite{aotgan} & \textbf{0.9100} & \textbf{28.878} & \textbf{34.550} \\
    \bottomrule
    \end{tabular}
     \vspace{-8pt}
    \label{tab:compare_models}
\end{table}

\subsection{Selecting EdgeConnect over AOTGAN}
\begin{figure*}
    \centering
    \setlength{\tabcolsep}{1pt}
    \newcommand{\imgwidth}{0.15\textwidth}
    \begin{tabular}{p{0.4cm}ccccccc}

        \multirow{-8}{*}{(a)} &
        \includegraphics[width=\imgwidth]{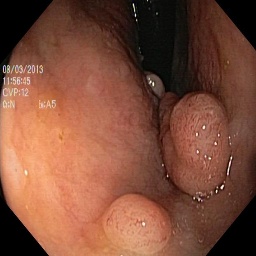} &
        \includegraphics[width=\imgwidth]{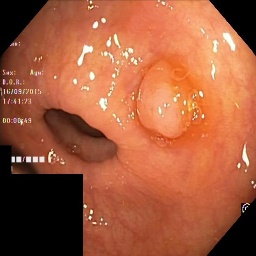} &
        \includegraphics[width=\imgwidth]{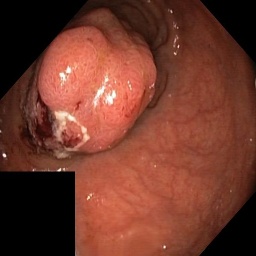} &
        \includegraphics[width=\imgwidth]{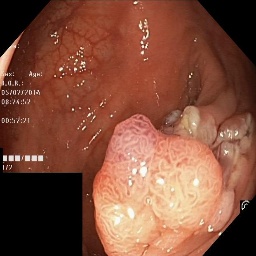} &
        \includegraphics[width=\imgwidth]{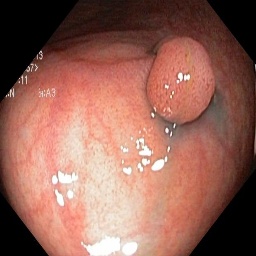} &
        \includegraphics[width=\imgwidth]{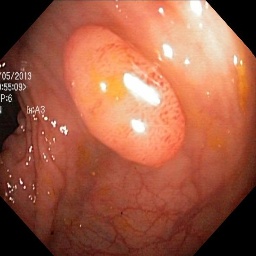}
        \\
        
        \multirow{-8}{*}{(b)} &
        \includegraphics[width=\imgwidth]{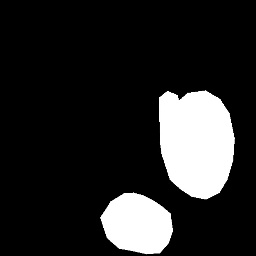} &
        \includegraphics[width=\imgwidth]{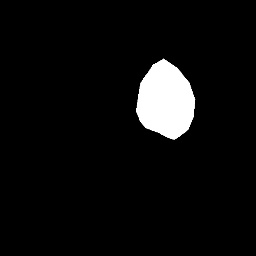} &
        \includegraphics[width=\imgwidth]{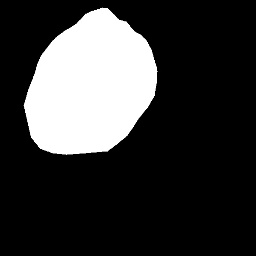} &
        \includegraphics[width=\imgwidth]{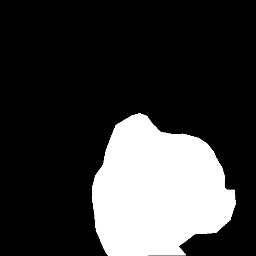} &
        \includegraphics[width=\imgwidth]{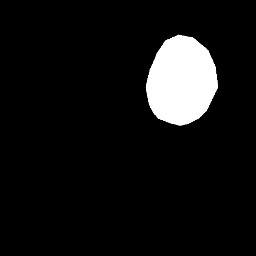} &
        \includegraphics[width=\imgwidth]{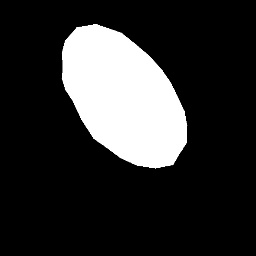}
        
        \\

        \multirow{-8}{*}{(c)} &
        \includegraphics[width=\imgwidth]{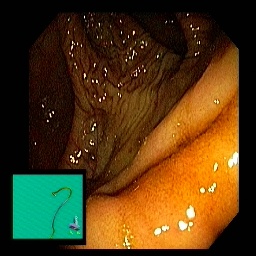} & 
        \includegraphics[width=\imgwidth]{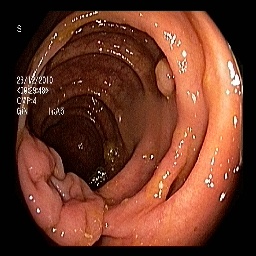} &
        \includegraphics[width=\imgwidth]{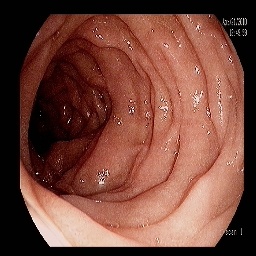} &
        \includegraphics[width=\imgwidth]{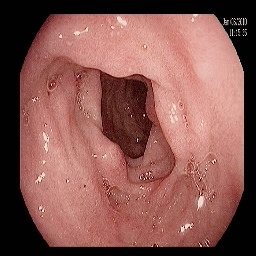} &
        \includegraphics[width=\imgwidth]{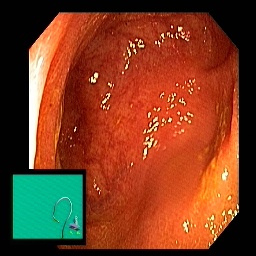} &
        \includegraphics[width=\imgwidth]{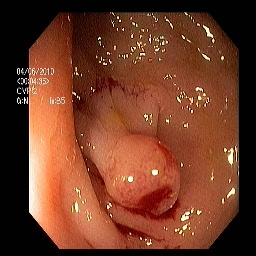}
        
        \\
        
        \multirow{-8}{*}{(d)} &
        \includegraphics[width=\imgwidth]{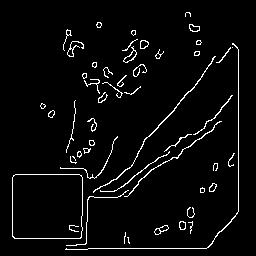} &
        \includegraphics[width=\imgwidth]{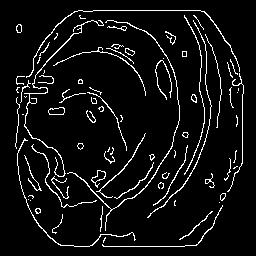} &
        \includegraphics[width=\imgwidth]{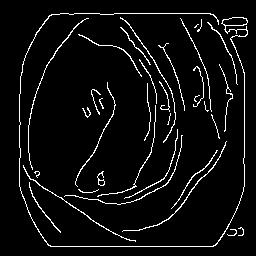} &
        \includegraphics[width=\imgwidth]{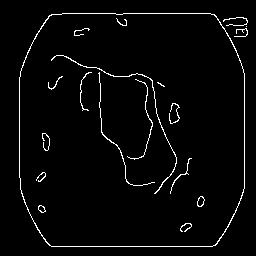} &
        \includegraphics[width=\imgwidth]{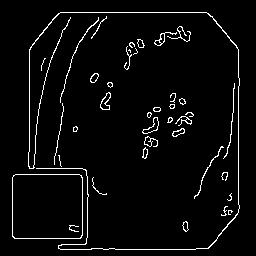} &
        \includegraphics[width=\imgwidth]{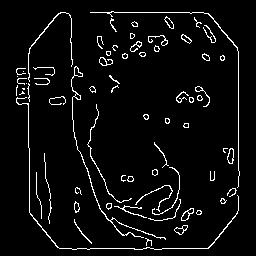} &
        
        \\
        
        \multirow{-8}{*}{(e)} &
        \includegraphics[width=\imgwidth]{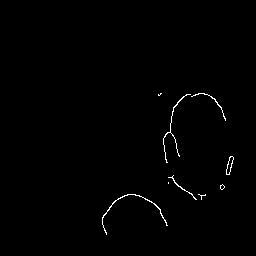} &
        \includegraphics[width=\imgwidth]{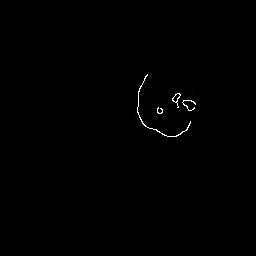} &
        \includegraphics[width=\imgwidth]{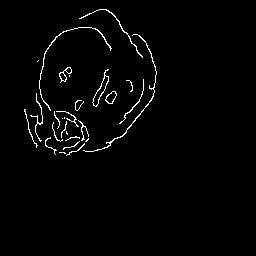} &
        \includegraphics[width=\imgwidth]{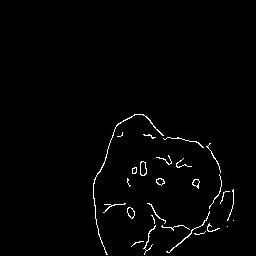} &
        \includegraphics[width=\imgwidth]{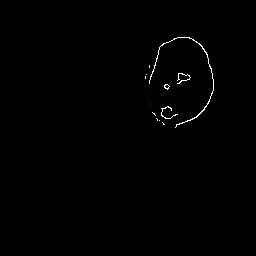} &
        \includegraphics[width=\imgwidth]{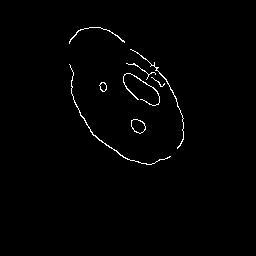} 
        
        \\

        \multirow{-8}{*}{(f)} &
         \includegraphics[width=\imgwidth]{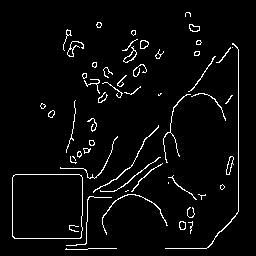} &
        \includegraphics[width=\imgwidth]{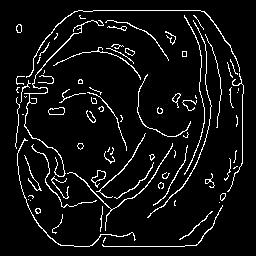} &
        \includegraphics[width=\imgwidth]{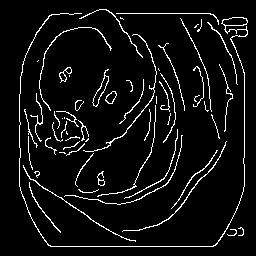} &
        \includegraphics[width=\imgwidth]{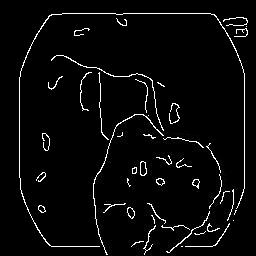} &
        \includegraphics[width=\imgwidth]{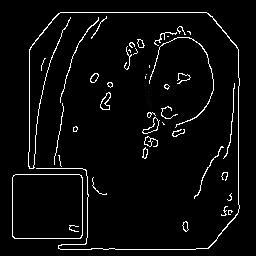} &
        \includegraphics[width=\imgwidth]{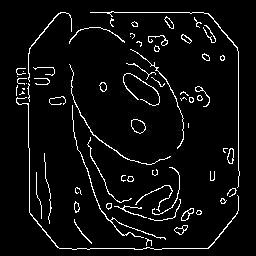} 
        
        \\
        
        \multirow{-8}{*}{(g)} &
        \includegraphics[width=\imgwidth]{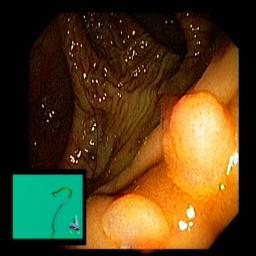} &
        \includegraphics[width=\imgwidth]{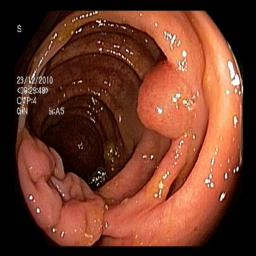} &
        \includegraphics[width=\imgwidth]{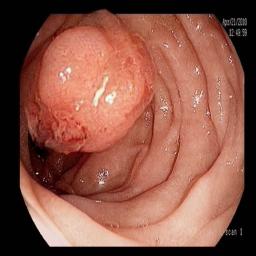} &
        \includegraphics[width=\imgwidth]{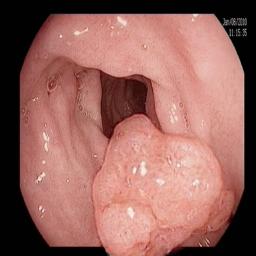} &
        \includegraphics[width=\imgwidth]{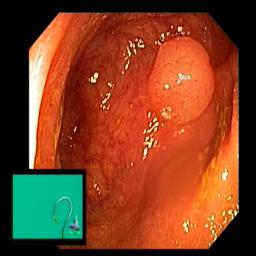} &
        \includegraphics[width=\imgwidth]{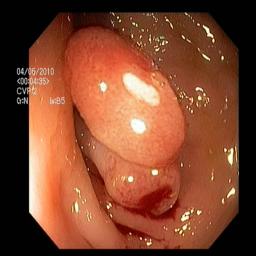}
         \\
         
         \multirow{-8}{*}{(h)} &
         \includegraphics[width=\imgwidth]{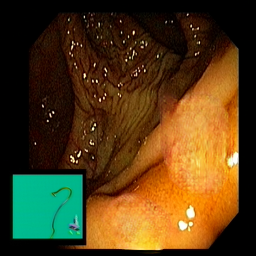} &
        \includegraphics[width=\imgwidth]{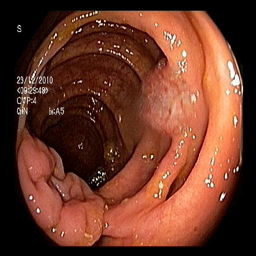} &
        \includegraphics[width=\imgwidth]{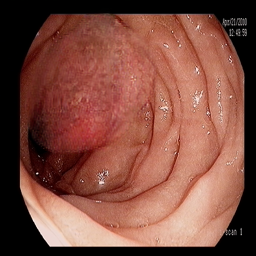} &
        \includegraphics[width=\imgwidth]{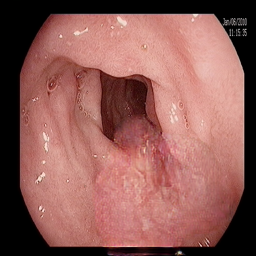} &
        \includegraphics[width=\imgwidth]{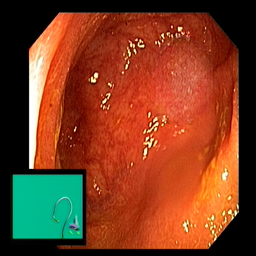} &
        \includegraphics[width=\imgwidth]{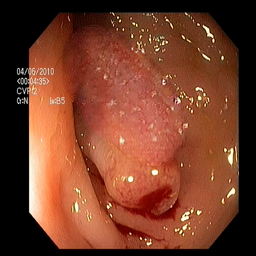}

    \end{tabular}
    
    \caption{Sample data used and generated in the different steps of PolypConnect pipeline. \textbf{(a)} - real polyp images, \textbf{(b)} - manually annotated polyp masks, \textbf{(c)} - randomly selected colon images used as input to the final step of PolypConnect, \textbf{(d)} - extracted edge images of row \textbf{c}. \textbf{(e)} - extracted edge images of polyp regions of row \textbf{a} using the masks of row \textbf{b}. \textbf{(f)} - combined edge images of row \textbf{d} and \textbf{f}. \textbf{(g)} - generated polyp on the images of row \textbf{c} using EdgeConnect. \textbf{(h)} - generated samples from AOTGAN.}
    \label{fig:gen_samples}
    \vspace{-20pt}
\end{figure*}

The performance metrics for Edgeconnect and AOTGAN on the validation data after fine-tuning the models, are shown in Table~\ref{tab:my_label}. In addition to qualitative evaluation, Figure~\ref{fig:gen_samples} provides example data from the different steps of the PolypConnect pipeline using the EdgeConnect model and the AOTGAN model. Due to obvious visual differences in the generated polyps between the models, we selected the EdgeConnect model as the main polyp inpainting model of the PolypConnect pipeline for further evaluation and qualitative assessment by domain experts.

\begin{table}[]
\centering
\caption{Calculated metrics for fine-tuned Edgeconnect and AOTGAN on the validation set for different fine-tune iterations.}

\begin{tabular}{|c| c| c| c| c| c| c|} 
 \toprule
 Model & \multicolumn{3}{|c|}{EdgeConnect} & \multicolumn{3}{|c|}{AOTGAN} \\
 \midrule
 Iteration & SSIM & PSNR & FID & SSIM & PSNR & FID\\ 
 \midrule

 500 & 0.529 & 17.832 & 77.712 & 0.882 & 27.114 & 27.114\\ 

 1000 & 0.527 & 17.859 & 77.007 & 0.890 & 28.176 & 42.102\\

 2000 & 0.527 & 17.847 & 76.460 & 0.887 & 28.021 & 42.449\\

 2500 & 0.526 & 17.836 & 76.956 & 0.889 & 27.969 & 42.484\\

 3000 & 0.527 & 17.817 & 77.461 & 0.888 & 28.045 & 42.154 \\ 

 3500 & 0.527 & 17.817 & 77.515 & 0.888 & 28.038 & 42.559\\  

 4000 & 0.528 & 17.832 & 76.988 & 0.889 & 28.058 & 41.628\\  

 4500 & 0.526 & 17.796 & 76.851 & 0.889 & 28.084 & 41.599\\  

 5000 &  0.525 & 17.860 & 77.219 & 0.889 & 28.100 & 42.200\\ 

 6000 & 0.527 & 17.835 & 76.310 & 0.882 & 28.0635 & 41.224\\  
 \bottomrule

\end{tabular}
%\vspace{-8pt}

\label{tab:my_label}
\end{table}

\subsection{Polyp segmentation models with synthetic polyps}
At this stage of the experiments, the generated polyps from PolypConnect (using the EdgeConnect model as the main inpainting model) are prepared for the segmentation evaluation. In total, there are four datasets. Therefore, we train four U-Net~\cite{ronneberger2015u} models for segmentation. The baseline dataset consists of only real polyp images. The remaining are datasets combining the real and generated polyp images. The first combined dataset consists of $800$ real and $800$ generated. The second and third are similar but with $1600$ and $2400$ generated polyp images. The models were evaluated on the same validation set of $200$ real images. The obtained metrics show a clear improvement in all models trained on the additional synthetic data. Results can be found in Table~\ref{tab:unetresults}.

\begin{table}

\caption{Evaluation of UNet segmentation model using real data and combined real and synthetic data. The best values are highlighted using \textbf{bold} text. Image IOU is calculated by aggregating intersection and union over whole dataset. Dataset IOU is also known as mIOU, and is the mean IOU.}
\begin{tabular}{|c |c |c |c |c |c|} 
 \toprule
 Dataset & Image IOU & Dataset IOU & Dice Coef & Prec & Rec \\ [0.5ex] 
 \midrule
 Baseline & 0.760 & 0.728 & 0.846 & 0.911 & 0.784 \\ 
 +800 & \textbf{0.795} & \textbf{0.765} & \textbf{0.874} & \textbf{0.923} & 0.817\\
 +1600 & 0.791 & 0.758 & 0.869 & 0.912 & \textbf{0.818} \\
 +2400 & \textbf{0.795} & 0.759 & 0.873 & 0.919 & 0.814 \\
 \bottomrule
\end{tabular}
\vspace{-8pt}
\label{tab:unetresults}
\end{table}
\begin{figure}
    \centering
    \setlength{\tabcolsep}{1pt}
    \newcommand{\imgwidth}{0.15\linewidth}
    \begin{tabular}{ccccccc}
   
    \includegraphics[width=\imgwidth]{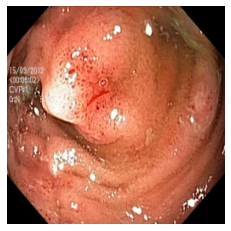}  &
    \includegraphics[width=\imgwidth]{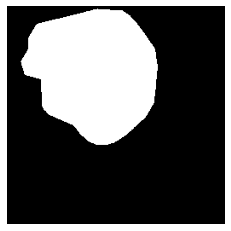}     &
    \includegraphics[width=\imgwidth]{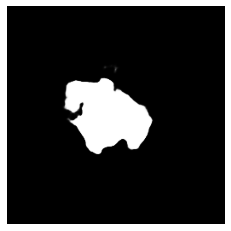}   &
    \includegraphics[width=\imgwidth]{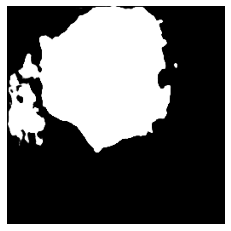}    &
    \includegraphics[width=\imgwidth]{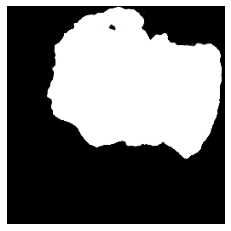}   &
    \includegraphics[width=\imgwidth]{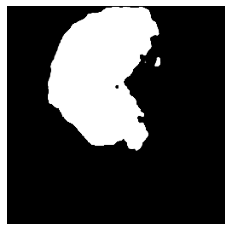}
    \vspace{-5pt}
    \\
    %\vspace{-5pt}
    
    \includegraphics[width=\imgwidth]{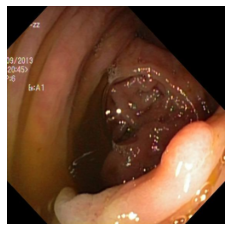}  &
    \includegraphics[width=\imgwidth]{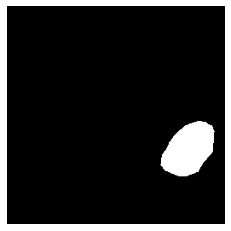}     &
    \includegraphics[width=\imgwidth]{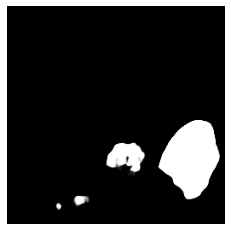}   &
    \includegraphics[width=\imgwidth]{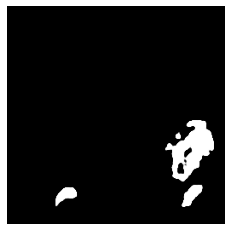}    &
    \includegraphics[width=\imgwidth]{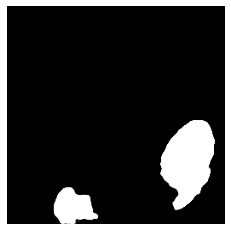}   &
    \includegraphics[width=\imgwidth]{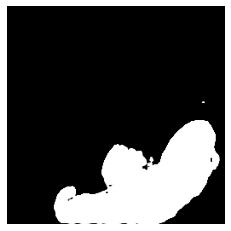}
    \vspace{-5pt}
    \\
    %\vspace{-5pt}
    
    \includegraphics[width=\imgwidth]{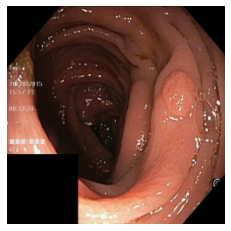}  &
    \includegraphics[width=\imgwidth]{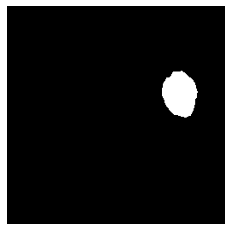}     &
    \includegraphics[width=\imgwidth]{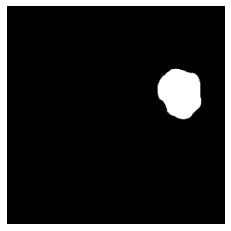}   &
    \includegraphics[width=\imgwidth]{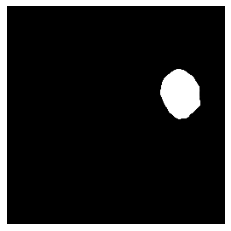}    &
    \includegraphics[width=\imgwidth]{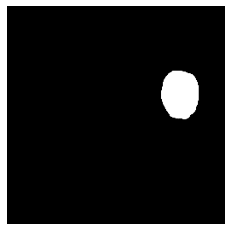}   &
    \includegraphics[width=\imgwidth]{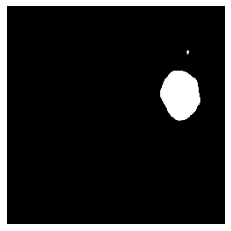}
    \vspace{-5pt}
    \\
    %\vspace{-5pt}
    
    \includegraphics[width=\imgwidth]{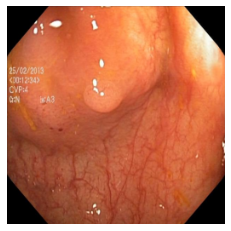}  &
    \includegraphics[width=\imgwidth]{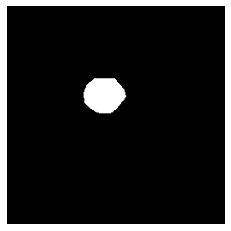}     &
    \includegraphics[width=\imgwidth]{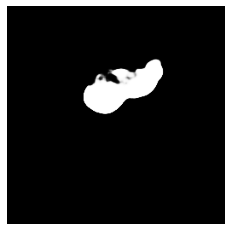}   &
    \includegraphics[width=\imgwidth]{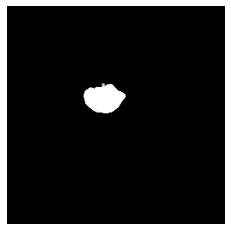}    &
    \includegraphics[width=\imgwidth]{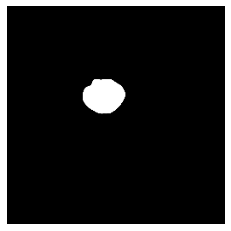}   &
    \includegraphics[width=\imgwidth]{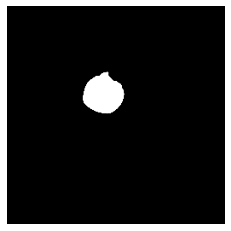}
    \vspace{-25pt}
    \\

    \\

    \\
    
    (a) & (b) & (c) & (d) & (e) & (f)

    \end{tabular}
    
\caption{Visual comparison of segmentation performance with UNet architecture when synthetic data is used. (a) Input Image. (b) Ground Truth. (c) Baseline (UNet) only with 800 real data (d) real data +800 (e) real data + 1600 (f) real data + 2400.}
\label{fig:seg_visual_comparison}
%\vspace{-12pt}
\end{figure}

\subsection{Qualitative analysis by domain experts}\label{sec:expert_survey}
Following the completion of the training and image inpainting of the generative models, a questionnaire was created to obtain subjective opinions on the generated images from domain experts. There were four participants in total from three fields of domain expert positions. Two of the participants are medical doctors (DOC), one is a gastroenterology consultant (GEC), and finally, an associate professor (SAP). The questionnaire included a total of ten polyp images and required the participants to rate images (fake or generated) on a confidence scale from $1-10$, where a score of $1$ means a real image, and a score of $10$ means a generated image. The summary of the collected results is tabulated in Table~\ref{tab:expert_response}. In addition, the participants were asked to give the same confidence rating only based on the polyps themselves, and the background surrounding the polyps. The participants were not given any information regarding the experiment and had no knowledge about the study. 

%\begin{table}[]
%    \centering
%    \caption{Qualitative evaluation of synthetic polyps over real polyps using an questionnaire.}
%    \begin{tabular}{|c|c|c|c|c|c|c|c|}
%        \toprule
%         Reader  & TP & FN & FP & TN & Accuracy & Recall & Precision  \\
%         \midrule
%         DOC  & 4 & 1 & 1 & 4 & 0.80 & 0.80 & 0.80 \\
%         DOC  & 3 & 2 & 3 & 2 & 0.50 & 0.60 & 0.50 \\
%         GEC  & 4 & 1 & 3 & 2 & 0.60 & 0.80 & 0.57 \\
%         SAP  & 3 & 2 & 3 & 2 & 0.70 & 0.80 & 0.66 \\
%         \midrule
%         Mean & - & - & - & -  & \textbf{0.65} & \textbf{0.75} & \textbf{0.65} \\
%         \bottomrule
%    \end{tabular}
%    \vspace{-10pt}
%    \label{tab:expert_response}
%\end{table}

\begin{table}[]
    \centering
    \caption{Qualitative evaluation of synthetic polyps over real polyps using an questionnaire.}
    \begin{tabular}{|c|c|c|c|c|c|c|c|}
        \toprule
         Reader  & TP & FN & FP & TN & Accuracy & Recall & Precision  \\
         \midrule
         DOC  & 4 & 1 & 1 & 4 & 80\% & 80\% & 80\% \\
         DOC  & 3 & 2 & 3 & 2  & 50\% & 60\% & 50\% \\
         SAP  & 3 & 2 & 3 & 2  & 70\% & 80\% & 66\% \\ 
         GEC  & 4 & 1 & 3 & 2  & 60\% & 80\% & 57\%\\ 
         GEC  & 3 & 2 & 1 & 4  & 70\% & 60\% & 75\% \\
         GEC  & 3 & 2 & 3 & 2  & 50\% & 60\% & 50\% \\
         GEC  & 3 & 2 & 5 & 0  & 30\% & 60\% & 37.5\% \\
         \midrule
         
         Mean & - & - & - & -  & \textbf{58.5\%} & \textbf{68.5\%} & \textbf{59.3\%} \\ 
         \bottomrule
    \end{tabular}
    \vspace{-10pt}
    \label{tab:expert_response}
\end{table}

\section{Discussion and conclusion}
\label{sec:discussion}

This study is focused on solving or reducing the data deficiency issues by efficiently generating realistic polyps in non-polyps images. This way, a dataset of finished segmented polyps can be generated in a short amount of time and vastly increase the data basis for polyp detection models. However, to be a useful solution, the generated results are required to be realistic and also improve the detection models experimentally.
Our idea incorporated utilizing non-segmented and unlabeled data for pretraining the models on general GI-tract image inpainting prior to finetuning for polyp generation. We used  ProGAN~\cite{karras2017progressive} to generate synthetic realistic polyp masks to be paired with random unlabeled images while training. However, we used the manually segmented polyp masks from KvasirSEG\cite{jha2020kvasir} to generate the synthetic polyps in non-polyp images. The polyp edges were extracted from the same images and used as the input for EdgeConnect. Synthetically generated pairs of polyp mask- and edge-images could be easily be created with ProGAN~\cite{karras2017progressive} or similar architectures, but this was not tested in this research. 
After conducting the experiments, we were able to generate realistic polyps in non-polyp images and also improve the detection rate of a polyp segmentation model by adding the synthetic data to the training data. The improved metrics are presented in Table~\ref{tab:unetresults}. Precision and recall increased by $1.2\%$ and $3.4\%$, respectively. Image IoU and dataset IoU increased from $0.76$ to $0.795$ ($4.6\%$) and $0.728$ to $0.765$ ($5.1\%$), respectively. The dice coefficient also showed improved results in the mixed datasets by $3.3\%$. All of the mixed datasets (utilizing the synthetic data) expressed clear improvements to the baseline. The model trained on the $+800$ dataset produced the overall best results. The $+1600$ and $+2400$ datasets yielded no clear improvement compared to the $+800$ dataset, and might therefore be an indication that additional synthetic images will not improve the segmentation model further. Moreover, the low accuracy of synthetic polyp detection by the domain experts presented in the results of the questionnaire implies that generated synthetic polyps are visually realistic as well.  

\section{Future works}

The PolypConnect pipeline can be enhanced by adding more pre-extracted features, such as Histogram of Oriented Gradients and texture features, etc. Furthermore, different \gls{gan} architectures can be experimented with to generate synthetic polyp masks, synthetic edge images, and synthetic clean colon images as well. Investigating to control more fine features of generated data can add value to the pipeline.

\section*{Acknowledgment}
The research presented in this paper has benefited from the Experimental Infrastructure for Exploration of Exascale Computing (eX3), which is financially supported by the Research Council of Norway under contract 270053.

\bibliographystyle{IEEEtran}
\bibliography{main}

\end{document}